\documentclass[lettersize,journal]{IEEEtran}
\usepackage{amsmath,amssymb,amsfonts}

\usepackage{algorithm}

\usepackage{array}
\usepackage[caption=false,font=normalsize,labelfont=sf,textfont=sf]{subfig}
\usepackage{stfloats}
\usepackage{url}
\usepackage{verbatim}
\usepackage{graphicx}
\usepackage{cite}
\usepackage{epsfig}
\usepackage{winsnotation}
\usepackage{colortbl}
\usepackage{xcolor}
\usepackage{graphicx}

\usepackage{bm}

\usepackage[noend]{algpseudocode}

\colorlet{shadecolor}{yellow}

\newcommand{\diag}{\mathop{\mathrm{diag}}}
\newcommand{\SVD}{\mathop{\mathrm{SVD}}}
\newcommand{\sel}{\mathop{\mathrm{sel}}}
\graphicspath{{../pdf/}{../jpeg/}}
\DeclareGraphicsExtensions{.pdf,.jpeg,.png}
\usepackage{url}
\usepackage{eqparbox}

\begin{document}
    \title{
    Beyond Traditional Beamforming: Singular Vector Projection for MU-MIMO}
  \author{Md Saheed~Ullah$^{*}$,
      Rafid Umayer~Murshed$^{*}$, Mohammad Saquib, and~Md. Forkan~Uddin 

\thanks{M.S. Ullah is with the Dept. of Electrical and Electronic Engineering (EEE), Bangladesh University of Engineering and Technology (BUET), Bangladesh and the Dept. of Electrical and Computer Engineering (ECE), University of Delaware, USA; R.U. Murshed is with the Dept. of EEE, BUET, Bangladesh and the Dept. of ECE, University of Texas at Dallas, USA;   M. Saquib is with the Dept. of ECE, University of Texas at Dallas, USA; M.F. Uddin is with the Dept. of EEE, BUET, Bangladesh. (e-mail:saheed@udel.edu; rafidumayer.murshed@utdallas.edu; saquib@utdallas.edu; mforkanuddin@eee.buet.ac.bd)}
\thanks{$^{*}$Denotes equal contribution (Their names appear in alphabetical order)}
      
\thanks{This research is supported by the ``Basic Research Grant" of Bangladesh University of Engineering and Technology.}

}

\maketitle

\begin{abstract}

This letter introduces a low-complexity beamforming approach for MU-MIMO systems with multiple data streams per user, minimizing inter-user interference and improving spectral efficiency (SE). The Interference-Optimized Singular Vector Beamforming (IOSVB) algorithm is developed by correlating inter-user interference with channel singular vectors. It blends interference minimization and SE maximization by identifying ideal singular vectors. Extensive simulations demonstrate that IOSVB provides near-optimal SE performance, closely matching exhaustive search results while reducing the computational overhead. This novel approach in MU-MIMO systems is a promising option for future 6G wireless communication networks due to its excellent performance and reduced complexity.

\end{abstract}

\begin{IEEEkeywords}
MU-MIMO, Beamforming, Interference Management, Spectral Efficiency, 6G Communications.
\end{IEEEkeywords}

\IEEEpeerreviewmaketitle


\section{Introduction}

\IEEEPARstart{T}{he} realm of wireless communication is poised for a revolutionary leap with the advent of 6G, where Multi-User Multiple Input Multiple Output (MU-MIMO) systems emerge as pivotal. MU-MIMO systems are at the forefront by harnessing an intricate array of antennas at both transmitters and receivers, promising to meet the escalating demand for higher data rates and network capacity \cite{intro_1_MIMO}. Integral to this evolution is the adoption of spatial multiplexing. This technique amplifies data transmission efficiency across the spatially distributed user spectrum, thereby marking a significant stride towards achieving the lofty throughput aspirations of 6G \cite{intro_SM}.

However, integrating spatial multiplexing within MU-MIMO systems is challenging, with effective interference management \cite{recent_research_MU_MIMO}. The complexity of effectively managing interference, particularly in environments where beam selection becomes crucial, cannot be overstated. As systems navigate the complexities of spectral efficiency (SE) and capacity maximization, the precision in beamforming vector selection becomes imperative. This precision directly influences the mitigation of inter-user interference and the enhancement of data throughput, thus emphasizing the crucial need for breakthroughs in beam-selection methodologies \cite{Inter_user_interference}.

In reviewing MU-MIMO beamforming optimization, it is clear that traditional methods, such as fractional programming and Hungarian algorithms, struggle with managing multiple data streams per user. Fractional programming, while effective for power control, falls short when applied to the complex spatial multiplexing of MU-MIMO, as it does not adequately address dynamic inter-user interference \cite{shen2018fractional}. Similarly, the Hungarian method, designed for assignment problems, cannot easily handle the interdependencies between beam optimizations, making it unsuitable for MU-MIMO beamforming where interference between beams is dynamic \cite{khan2020optimizing,al2022low}. DFT codebook-based approaches, although efficient in practical systems, are limited by static, predefined beam patterns that cannot adapt to instantaneous channel conditions\cite{penna2017search}. These methods are unable to fully optimize beamforming in real-time, reducing their effectiveness in complex, high-interference environments like MU-MIMO. In light of the limitations highlighted, there is a clear need for novel approaches that effectively manage interference while supporting multiple data streams per user.

Motivated by the identified challenges and limitations, this study introduces the Interference-Optimized Singular Vector Beamforming (IOSVB) algorithm, a reduced-complexity solution for interference management in MU-MIMO systems. Traditional algorithms \cite{yuan2018hybrid}, which iteratively optimize the rate or directly manage inter-user interference, are computationally intensive and impractical for real-time applications. In contrast, IOSVB leverages an upper bound of total system interference as a computationally efficient metric for iterative optimization. Notably, the combiner at the user end requires no knowledge of other users' channels. This strategic design centralizes the computational burden at the base station (BS), ensuring minimal load on user devices with limited computational resources. Empirical results demonstrate that IOSVB achieves near-optimal SE with significantly reduced complexity, providing a feasible solution for MU-MIMO. 

\emph{Notation}: $\RM{X}[\RS{\mathcal{I}}]$ denotes selecting columns from matrix $\RM{X}$ indexed by set $\RS{\mathcal{I}}$, and $\RM{X}[\RS{\mathcal{I}},\RS{\mathcal{I}}]$ selects both columns and rows. $\RV{X^\dag}$ and $\RM{X}^\dag[\RS{\mathcal{I}}]$ represent the conjugate transpose of $\RV{X}$ and its indexed selection, respectively. $||.||$ denotes Frobenius norm.

\newcommand{\ith}[1]    {{#1}^{\underline{ \text{th}}}}

\section{System Model and Problem Formulation}

We consider a downlink MU-MIMO system consisting of a BS with \( N_t \) transmit antennas and \( K \) users, each with \( N_r \) receive antennas, where each user is served with \( N_s \) data streams. This work focuses on an unconstrained fully digital beamformer design using a beam assignment strategy for both transmitter and receiver. The fully digital beamforming matrices can later extract practically implementable hybrid beamforming matrices \cite{self_HBF}. The transmitted signal \( \RV{x}_k \) is obtained through the digital beamforming matrix \( \RM{F}_k \in \mathbb{C}^{N_t \times N_s}\) and the data symbol vector \( \RV{s}_k \in \mathbb{C}^{N_s \times 1} \). The channel matrix for the \( \ith{k} \) user is $\RM{H}_k\in \mathbb{C}^{N_r \times N_t}$. Moreover, $\RM{W}_k \in \mathbb{C}^{N_r \times N_s}$ represents the receive beamforming matrix for the $\ith{k}$ user. The SE for the $\ith{k}$ user is
\begin{equation} 
\label{eq_rate}
R_k = \log_2 \left| \RM{I}_{N_r} + \RM{W}_{k} \RM{C}_{k}^{-1} \RM{W}_{k}^\mathrm{\dag} \RM{H}_{k} \RM{F}_{k} \RM{F}_{k}^\mathrm{\dag} \RM{H}_{k}^\mathrm{\dag}\right|,
\end{equation}
where,
\begin{equation}
\RM{C}_{k} = \RM{\Delta}_{k} + N_0^2 \RM{W}_{k}^\mathrm{\dag} \RM{W}_{k},
\end{equation}
with noise power $N_0$ and the inter-user interference matrix 
\begin{equation} \label{eq_interference}
\RM{\Delta}_{k} = \RM{W}_{k} \; \RM{H}_{k} \left( \sum_{j \neq k} \RM{F}_{j} \RM{F}_{j}^\dag \right) \RM{H}_{k}^\mathrm{\dag} \RM{W}_{k}.
\end{equation}
The primary challenge is to find the precoder $\RM{F}_k$ and combiner $\RM{W}_k$ that minimize interference to maximize the sum rate. 


\textbf{Channel Model}: The performance of the proposed approach is evaluated using the NYUSIM model for sub-THz bands. The NYUSIM model simulates sub-THz band complexities, particularly in indoor scenarios \cite{ju2023142}. The NYUSIM channel simulator can produce an accurate 3-dimensional angular power spectrum, power delay profiles (PDPs), and omnidirectional and directional channel impulse responses (CIRs) at sub-THz\cite{ju2019millimeter}.

\section{Proposed Beamforming Approach}

Here, we introduce the IOSVB algorithm to iteratively yield near-optimal beamforming matrices, $\RM{F}_k$ and $\RM{W}_k$. IOSVB uses channel singular vector correlations to reduce inter-user interference to select the most effective vectors within candidate matrices $\RM{\tilde{F}}$. Since our system iteratively searches, we need a metric to identify the best beamforming vectors. The rate equation (\ref{eq_rate}) or system interference (\ref{eq_interference}) would be too computationally intensive for this iterative approach. This leads us to derive an alternate representation of the interference, an upper bound of the total system interference. 

The algorithm iteratively selects a subset of optimized categorical column indices, $\mathcal{I}_k$, for each user $k$, containing $N_s$ values corresponding to $N_s$ beamforming directions. Each element of the set $\mathcal{I}_k$ uniquely identifies a particular beam-direction. The aggregation of these subsets across all $K$ users forms the global index set $\mathcal{I}$, defined as  $\mathcal{I} = \bigcup_{k=1}^{K} \mathcal{I}_k$. The relationship $\RM{\tilde{F}}[\mathcal{I}] = \RM{F}$ succinctly captures the key essence of our approach, signifying that the optimal matrices $\RM{F}= [\RM{F}_1,\cdots,\RM{F}_k,\cdots,\RM{F}_K]
$ are a strategic selection from the broader candidate set $\RM{\tilde{F}}$.

Upon performing SVD on the channel matrix, two orthonormal matrices $\RM{U}_k\in\mathbb{C}^{N_r \times N_r}$, $\RM{V}_k\in \mathbb{C}^{N_t \times N_t}$ and a diagonal matrix $\RM{S}_k\in \mathbb{R}^{N_r \times N_t}$, i.e. $\SVD(\RM{H}_k)=\RM{U}_k\RM{S}_k\RM{V}_k^\dag$. Given the sparse nature of channels at mmWave, utilizing a search space of all $N_t$ vectors results in extensive computational complexity at the expense of minimal performance gain. Thus, we narrow down the search domain by discarding vectors in $\RM{V}_k$ associated with minimal channel gain. We select only the first $N_c$ columns from $\RM{V}_k$, where $N_c$ is chosen to be close to the rank of the channel matrix. This selection is informed by the channel gains in $\RM{S}_k$ and compresses the search space to $N_{\text{ex}} = {\binom{N_c}{N_s}}^K$ possible combinations instead of ${\binom{N_t}{N_s}}^K$, considerably simplifying the optimization challenge. We derive $\RM{\tilde{F}}_k \in \mathbb{C}^{N_t \times N_c}$ by extracting the first $N_c$ columns from $\RM{V}_k$ for all the users. We also form $\RM{S}^\prime_k$ by selecting the first $N_c$ columns and first $KN_c$ rows from $\RM{S}_k$. 
The concatenated candidate matrix $\RM{\tilde{F}}= [\RM{\tilde{F}}_1,\cdots,\RM{\tilde{F}}_k,\cdots,\RM{\tilde{F}}_K]
$ represents the collation of selected matrices for all users.
Similarly, the aggregated singular vector matrix for all users, denoted as $\RM{S} = [\RM{S}^\prime_1,\cdots \RM{S}_k^\prime, \cdots, \RM{S}_K^\prime]$, spans a dimension of $K N_c \times K N_c$. This diagonal matrix $\RM{S}$ can also be represented as $\diag{(\sigma_1,\sigma_2,\cdots,\sigma_{KN_c})}$. The correlation matrix for all combinatorial selections, symbolized by $\RM{\Lambda}_c$, is formulated as
\begin{equation}
\label{csel}
    \RM{\Lambda}_c = {\RM{S}}\;\RM{\tilde{F}}^\dag \; \RM{\tilde{F}},
\end{equation}
where, $\RM{\Lambda}_c$ is a  $KN_c \times KN_c$ matrix. From $\RM{\Lambda}_c$, we generate $N_{\text{ex}}$ combinations of matrices, each with dimensions $K N_s \times K N_s$. A particular combination matrix within this set is represented by $\RM{\Lambda}$, expressed as 
 \begin{equation}\label{lmbda}
     \RM{\Lambda} = \RM{S}[\RS{\mathcal{I}},\RS{\mathcal{I}}]\;{\RM{\tilde{F}}}^\dag[\RS{\mathcal{I}}] \; \RM{\tilde{F}}[\RS{\mathcal{I}}].
      \end{equation}
Note that $\RM{\Lambda}$ is a $KN_s \times KN_s$ square matrix that can be extracted from $\RM{\Lambda}_c$ using $\RM{\Lambda} = \RM{\Lambda}_c[\RS{\mathcal{I}},\RS{\mathcal{I}}]$.

From (\ref{eq_interference}), the interference of the system can be written as, 
\begin{equation} \label{eq_6}
||\RM{\Delta}_{k}|| = ||\RM{W}_{k}^\mathrm{\dag} \RM{H}_{k} \left( \sum_{j \neq k} \RM{F}_{j} \RM{F}_{j}^{\dag} \right) \RM{H}_{k}^\mathrm{\dag} \RM{W}_{k}|| .
\end{equation}
Lemma 1, proved in the Appendix, allows us to rewrite (\ref{eq_6}) as 
\begin{equation} \label{eq_7}
||\RM{\Delta}_{k}|| = ||\sum_{j \neq k}
(\RM{W}_{k}^\mathrm{\dag} \RM{H}_{k} \RM{F}_{j})
({\RM{W}_{k}^\mathrm{\dag} \RM{H}_{k} \RM{F}_{j}})^\mathrm{\dag}||  ,
\end{equation}
and then triangle inequality is used to obtain 
\begin{equation} \label{eq_8}
||\RM{\Delta}_{k}|| \leq \sum_{j \neq k}
{||\RM{W}_{k}^\mathrm{\dag} \RM{H}_{k} \RM{F}_{j}||}^2 .
\end{equation}
Lemma 2, proved in the Appendix, yields 
\begin{equation} \label{eq_9}
   ||\RM{\Delta}_k||\leq \sum_{j \neq k}{|| \RM{S}_k[\RS{\mathcal{I}_k},\RS{\mathcal{I}_k}] \; \RM{V}_k^\dag[\RS{\mathcal{I}_k}]  \;\RM{V}_j[\RS{\mathcal{I}}_j]||}^2 ,
\end{equation}
where, $\RM{S}_k[\RS{\mathcal{I}}_k,\RS{\mathcal{I}}_k]$ a matrix of span $N_s \times N_s$ that represents the down-selection of the columns and rows from $\RM{S}_k$ that correspond to the index set $\RS{\mathcal{I}}_k.$ 
If the total interference over all users is expressed as $\RM{\Delta}$, in other words, $\sum_{k=1}^K ||\RM{\Delta}_k|| = \RM{\Delta}$, then using (\ref{eq_9}), we have
\begin{equation} \label{eq_10}
   \RM{\Delta} \leq \sum_{k=1}^K \sum_{j \neq k}^K{|| \RM{S}_k[\RS{\mathcal{I}_k},\RS{\mathcal{I}_k}] \;\RM{V}_k^\dag[\RS{\mathcal{I}_k}]  \;\RM{V}_j[\RS{\mathcal{I}}_j]||}^2 .
\end{equation}
By definition, $\RM{\tilde{F}}_k[\RS{\mathcal{I}}_k] = \RM{V}_k[\RS{\mathcal{I}}_k]$. Thus, we can rewrite (\ref{eq_10}) as 

\begin{equation}\label{eq_11}
   \RM{\Delta} \leq \sum_{k=1}^K \sum_{j \neq k}^K{|| \RM{S}_k[\mathcal{I}_k,\mathcal{I}_k] \;\RM{\tilde{F}}_k^\dag[\RS{\mathcal{I}}_k]  \;\RM{\tilde{F}}_j[\RS{\mathcal{I}}_j]||}^2.
\end{equation}

Noting that $\RM{\tilde{F}}$ is the concatenation of all the $\RM{\tilde{F}}_k$ matrices, and $\mathcal{I}_k$ indicates the beam directions for the $\ith{k}$ user, we can extract $\RM{\tilde{F}}_k$ from $\RM{\tilde{F}}$ using the directions $\mathcal{I}_k$, such that \(\RM{\tilde{F}}_k[\RS{\mathcal{I}}_k] = \RM{\tilde{F}}[\RS{\mathcal{I}}_k]\) and similarly, \(\RM{S}_k[\RS{\mathcal{I}}_k,\RS{\mathcal{I}}_k] = \RM{S}[\RS{\mathcal{I}}_k,\RS{\mathcal{I}}_k]\). This allows us to rewrite (\ref{eq_11}) as

\begin{equation}\label{eq12}
   \RM{\Delta} \leq \sum_{k=1}^K \sum_{j \neq k}^K{|| \RM{S}[\mathcal{I}_k,\mathcal{I}_k] \;\RM{\tilde{F}}^\dag[\RS{\mathcal{I}}_k]  \;\RM{\tilde{F}}[\RS{\mathcal{I}}_j]||}^2.
\end{equation}

For the correlation between the beam directions of the $\ith{k}$ and $\ith{j}$ users, the $\RM{\Lambda}$ matrix in (\ref{lmbda}) can be refined as \(\RM{\Lambda}[\RS{\mathcal{I}}_k,\RS{\mathcal{I}}_j] = \RM{S}[\RS{\mathcal{I}}_k,\RS{\mathcal{I}}_k] \; {\RM{\tilde{F}}}^\dag[\RS{\mathcal{I}}_k] \; \RM{\tilde{F}}[\RS{\mathcal{I}}_j]\). This refinement simplifies (\ref{eq12}) to

\begin{equation}\label{eq_13}
   \RM{\Delta} \leq \sum_{k=1}^K \sum_{j \neq k}^K{|| \Lambda [\RS{\mathcal{I}}_k,\RS{\mathcal{I}}_j]||}^2.
\end{equation}

Clearly, $\RM{\Lambda}[\RS{\mathcal{I}}_k,\RS{\mathcal{I}}_j]$ is a sub-matrix of $\RM{\Lambda}$, containing $N_s \times N_s$ elements. Considering all such sub-matrices for all $\RS{\mathcal{I}}_k$ and $\RS{\mathcal{I}}_j$, we have \(||\RM{\Lambda}||^2 = \sum_{k=1}^K \sum_{j=1}^K{|| \RM{\Lambda}[\mathcal{I}_k,\mathcal{I}_j]}||^2\). The constraint \(j \neq k\) in (\ref{eq_13}) removes the diagonal components from the norm calculation. Consequently, we can rewrite (\ref{eq_13}) to bound the total interference as

\begin{equation}\label{optim}
    \RM{\Delta} \leq {||\RM{\Lambda} - \diag (\RM{\Lambda})||}^2.
\end{equation}

\begin{figure}
  \begin{center}
  \includegraphics[height=4cm,trim={1.4cm 1cm 2cm 1.9cm},clip]{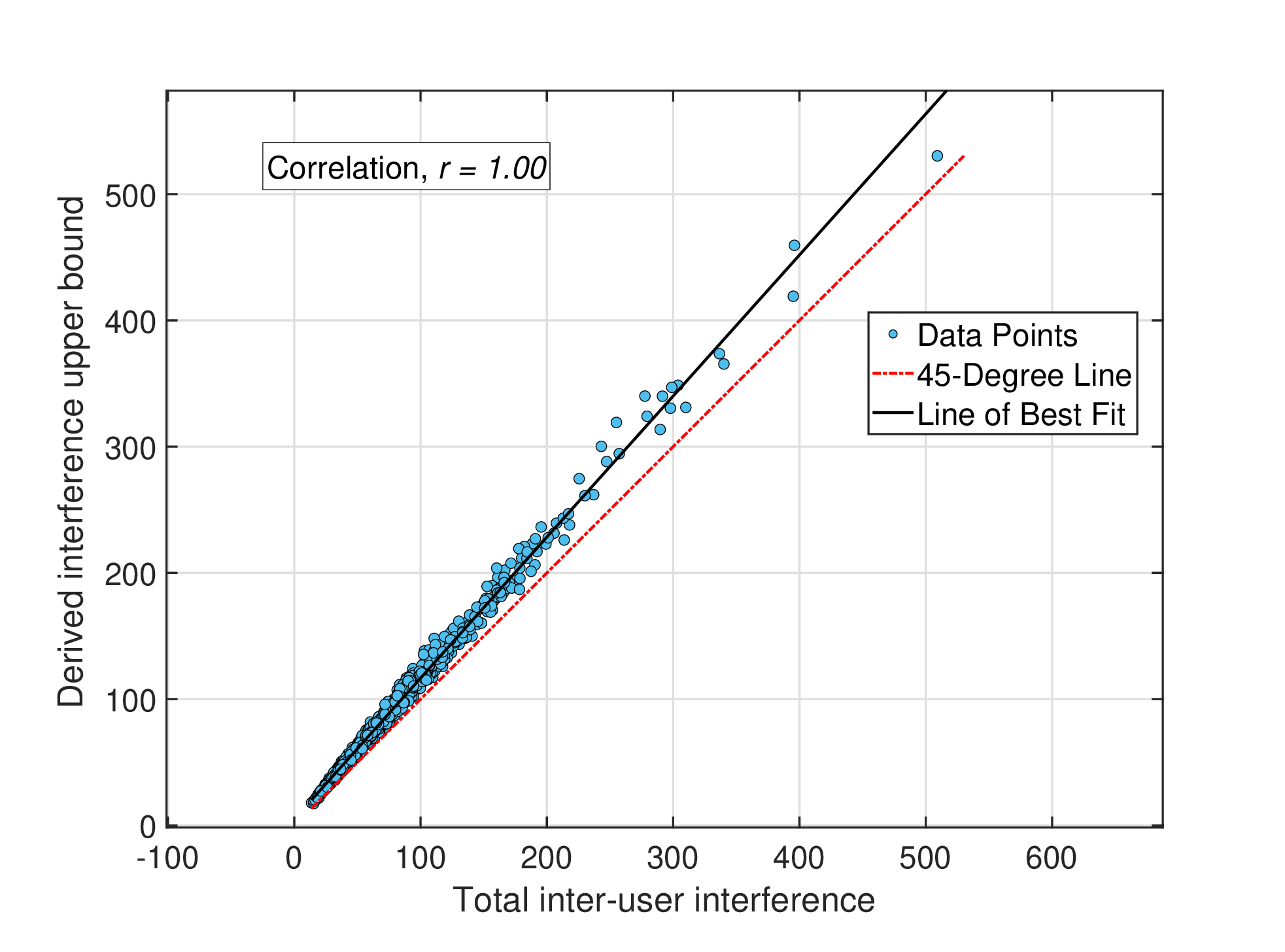}
  \caption{{Scatter plot of actual system interference and the upper bound in (\ref{optim}).}}
  \label{fig_proof}
  \end{center}
\end{figure}

\begin{algorithm}
	\caption{Proposed IOSVB}
	\label{iosvb}
	\begin{algorithmic}[1]
	
 \State Input: 
  $\{\RM{H}_k\}_{k=1}^K$, $N_s$, $N_c$, $\gamma$, $\rv{f}(0)=10^9$
  \State Calculate $\RM{U}_k, \RM{S}_k, \RM{V}_k^\dag$ from $\RM{H}_k$ and find $\RM{\tilde{F}}_k$ from $\RM{V}_k$;
  \State Calculate $\sigma_{\max}$ using $\sigma_{\max} =\sum_{j=1}^{K}\sum_{i=1}^{N_s}\sigma_{j,i}$;
 \State Calculate $\RM{\tilde{F}}$ by concatenating all $\RM{\tilde{F}}_k$;
       \State Calculate $\RM{\Lambda}_c$ using (\ref{csel});
       \State For $\{{\RS{\mathcal{I}}^i}\}_{i=1}^{N_{\text{ex}}}$ sets of indices, calculate $\RM{\Lambda}^i = \RM{\Lambda}_c[\RS{\mathcal{I}}^i,\RS{\mathcal{I}}^i]$;
       \State Set $i=1$, $i_{\min}=1$;
 
	\Repeat : if ($\sigma_{\sel} > \gamma*\sigma_{\max}$)
    \State Find $\RM{\Lambda}^i$ for $\ith{i}$ instance;
    \State  Calculate $\rv{f}(i)= {||\RM{\Lambda}^i-\diag(\RM{\Lambda}^i)||} $;
      \State If $\rv{f}(i)<\rv{f}(i_{\min})$; put $i_{\min}=i$	;
      \State Set $i=i+1$;
       \Until $i> N_{\text{ex}}$ 
       \State Find $\RM{F}_k = \RM{V}_k[\RS{\mathcal{I}}^{i_{\min}}_k]$, $\RM{W}_k = \RM{U}_k[\RS{\mathcal{I}}^{i_{\min}}_k]$ 
	\end{algorithmic}
\end{algorithm}

 Fig. \ref{fig_proof} presents a scatter plot comparing the upper bound and total interference from (\ref{optim}) using a thousand channel realizations. The plot shows a near-perfect correlation (\textit{r} = 1) between the total interference ($\RM{\Delta}$) and $||\RM{\Lambda} - \diag (\RM{\Lambda})||^2$, confirming the tightness of the upper bound. Moreover, the plot demonstrates that the total interference remains consistently bounded by $||\RM{\Lambda} - \diag (\RM{\Lambda})||^2$, even when user beam directions are highly correlated. Finally, the optimization turns into


\begin{subequations} \label{eq_optimize}
\begin{IEEEeqnarray}{RCL} 
    & \underset{\RM{\Lambda}}{\text{minimize}}
    & \qquad \rv{f} = ||\RM{\Lambda} - \diag(\RM{\Lambda})||  \label{eq:objective} \\*
    & \text{subject to}
    & \qquad \sigma_{\sel} > \gamma \sigma_{\max} \IEEEeqnarraynumspace \label{eq:constraint}
\end{IEEEeqnarray}
\label{cor}
\end{subequations}
where, $\gamma$ is an arbitrary channel gain threshold, $0<\gamma<1$, $\sigma_{\max}$ indicates the maximum possible aggregated channel gain given by  $\sigma_{\max} =\sum_{j=1}^{K}\sum_{i=1}^{N_s}\sigma_{j,i}$, and $\sigma_{\sel}$ is the selected channel gain sum defined as  $\sigma_{\sel}=\sum_{j=1}^{K}\sum_{i\in \RS{\mathcal{I}}_i}\sigma_{j,i}$. Here, (\ref{eq:constraint}) ensures that the selected beams are strong enough to contribute to the data rate. We postulate that minimizing the interference upper-bound through (\ref{eq_optimize}) indirectly maximizes the lower-bound of the sum-rate in (\ref{eq_rate}). The optimal precoder and combiner matrices $\RM{F}_k$ and $\RM{W}_k$ are then selected using the combination from $\RM{\Lambda}$ for which the minimum value of $\rv{f}$ is found in (\ref{cor}). Once $\RS{\mathcal{I}} = \RS{\mathcal{I}}^{i_{\min}}$ is obtained (Algorithm \ref{iosvb}), the optimal \(\RM{W}_k\) is extracted via \(\RM{W}_k = \RM{U}_k[\RS{\mathcal{I}}_k]\). Notably, each receiver only requires the final indices \(\RS{\mathcal{I}}_k\), allowing independent processing of data streams without knowledge of other users' channels. This formulation centralizes all computational burden at the BS, eliminating the need for user devices to run the IOSVB algorithm. The IOSVB algorithm can be seamlessly adapted to the uplink scenario, where the BS applies the IOSVB algorithm and finds optimal receive beamforming vectors.
\\
\textbf{Complexity}: The proposed IOSVB algorithm exhibits a complexity order of $\mathcal{O}(N_{\text{iter}} {(KN_s)}^2+K N_tN_r\min(N_t,N_r))$ including the complexity of the SVD, where $N_{\text{iter}}$ denotes the number of iterations required for optimization. In contrast, the exhaustive search algorithm employs brute force exploring the $N_c$ columns for all users to find the maximum SE in (\ref{eq_rate}), resulting in significantly higher complexity of $\mathcal{O}(N_{\text{ex}} {(KN_s)}^2)+ K N_tN_r\min(N_t,N_r))$. Note that $N_{\text{ex}} \hspace{-0.5mm} \geq \hspace{-0.5mm} N_{\text{iter}} \hspace{-0.5mm} \geq \hspace{-0.5mm} 1$, with $N_{\text{iter}}$ being influenced by the channel gain ratio $\gamma$. For large $\gamma$, $N_{\text{iter}}\ll N_{\text{ex}}$.

\section{Results}
In this section, we provide the simulation parameters, evaluate the proposed IOSVB algorithm's performance, and compare it with the performance of the existing alternatives.
  
\subsection{Simulation Setup and Parameters}
The simulation results were generated using a computer with a Core i7 processor and 32GB RAM. For the simulation, the transmit and receive ends are considered to be equipped with a planner array of 144 and 36 elements, respectively. The total number of users in the system is 5.

\subsection{Selection of Algorithmic Parameters}

\begin{figure}
  \begin{center}
\includegraphics[width=\columnwidth,height=4.80cm,trim={1.2cm 0cm 1cm 1cm},clip]{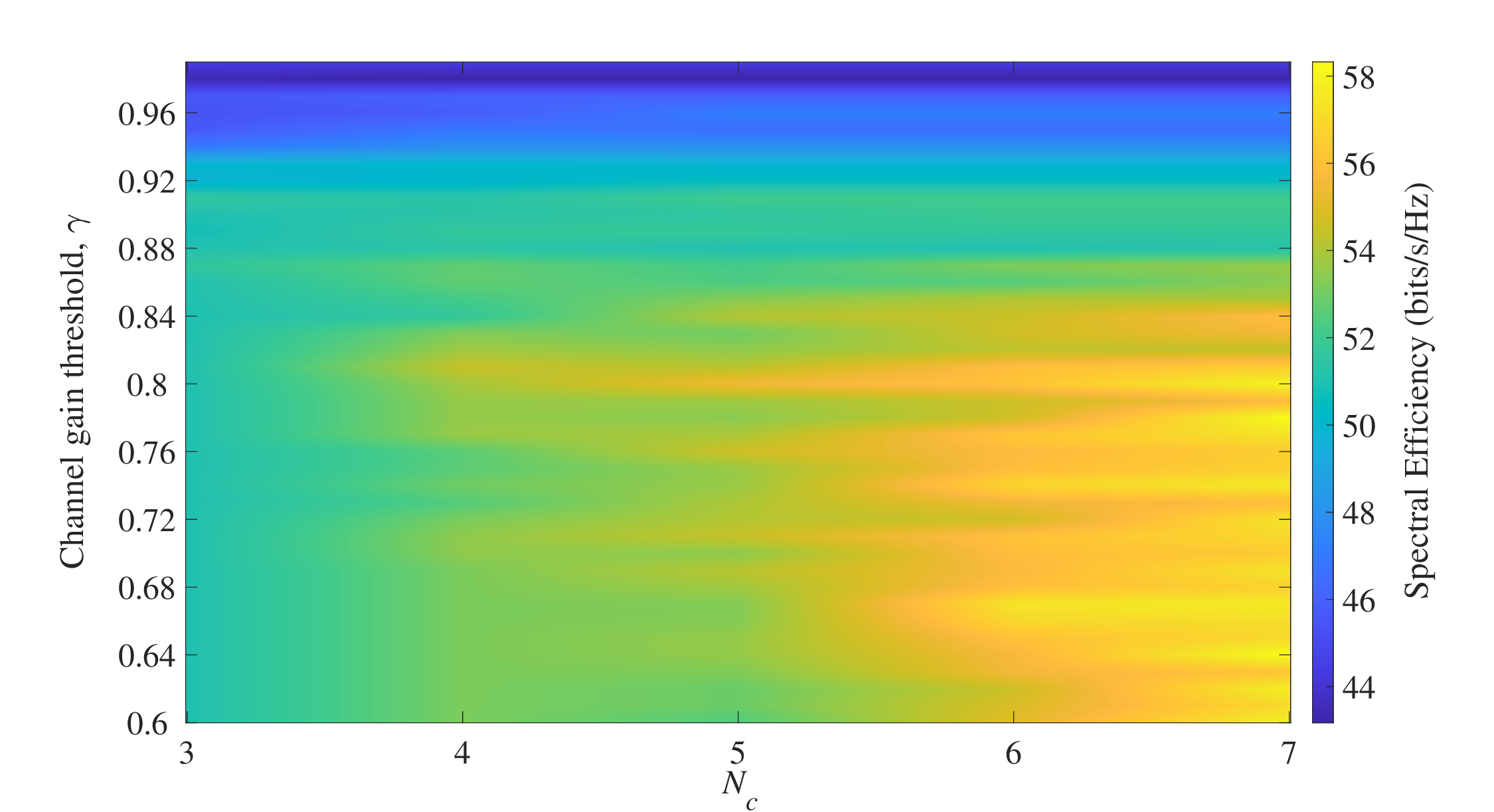}
  \caption{{SE with respect to candidate columns, $N_c$ and channel gain threshold, $\gamma$} for $N_s = 2.$ }
  \label{joint_selection_Rsel_Gamma}
  \end{center}
\end{figure}

\begin{figure}
  \begin{center}
  \includegraphics[width=\columnwidth,height=4.8cm,trim={0.8cm 0cm 0cm 1cm},clip]{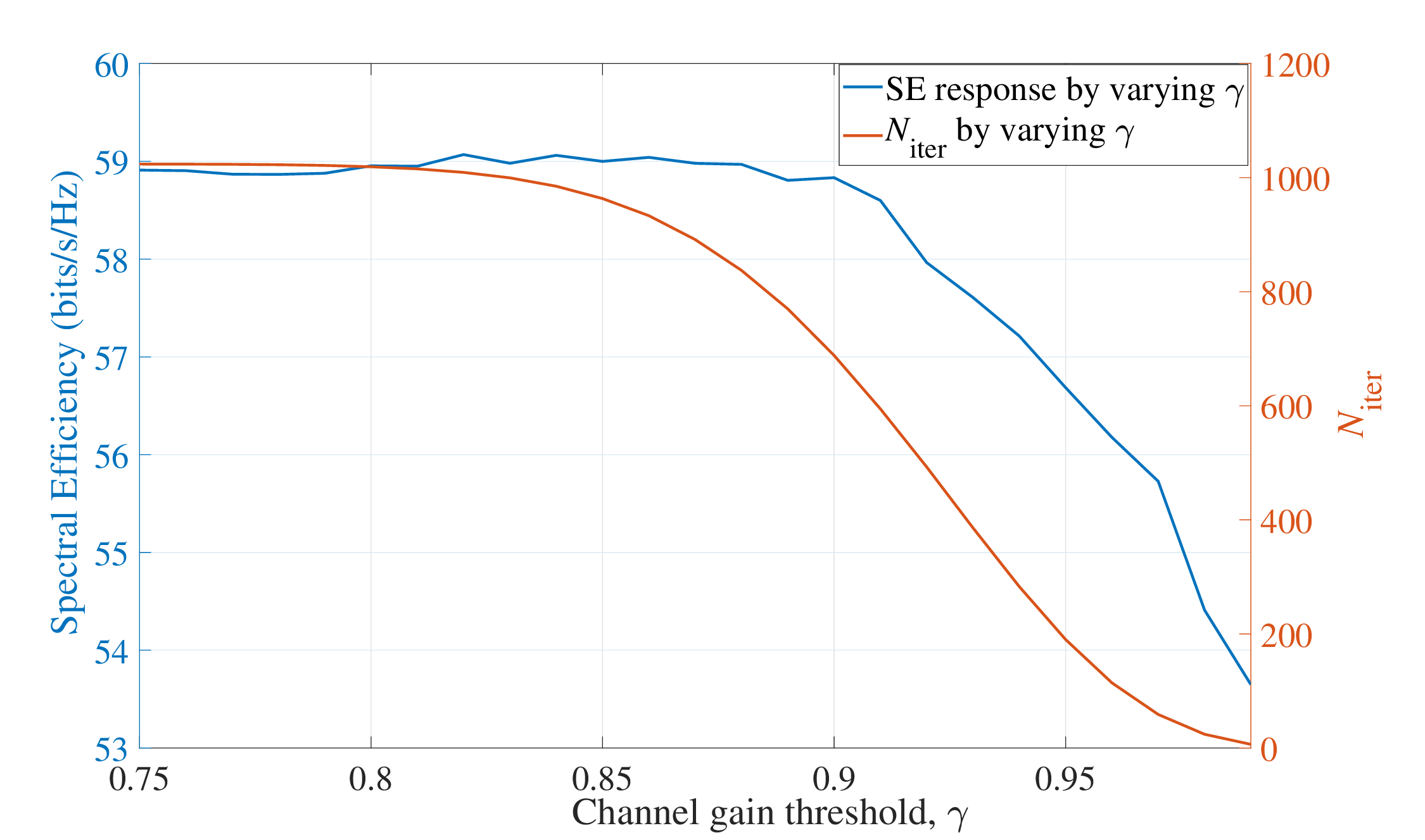}
  \caption{{SE and number of iterations $N_\text{iter}$ by varying channel gain ratio $\gamma$.}}
  \label{gamma_se_n}
  \end{center}
\end{figure}

We examine the impact of key algorithmic parameters (\(N_c\), \(N_{iter}\), and \(\gamma\)) on the SE of a MU-MIMO system using the IOSVB algorithm. Simulations over 1000 channel realizations with \(N_s = 2\) highlight the influence of these parameters on SE, as shown in Fig. \ref{joint_selection_Rsel_Gamma} and \ref{gamma_se_n}. 



An analysis of the IOSVB algorithm, illustrated in Fig. \ref{gamma_se_n}, indicates the relationship between spectral efficiency and computational cost in terms of $N_{iter}$ by varying $\gamma$. Fig. \ref{gamma_se_n} shows an inverse relationship between the number of iterations required and \(\gamma\). This is due to the reduced number of combination checks as \(\gamma\) increases, identifying an optimal \(\gamma\) value around $0.80$ for efficient algorithm performance. Larger values of \(\gamma\) result in fewer candidate beam directions, thereby reducing the search space and the number of iterations ($N_{\text{iter}}$) required for convergence. It can be demonstrated that the optimization of $\gamma$ exhibits a similar trend, even when Nc and Ns are varied.
Table \ref{selection_Rsel_Ns} identifies optimal \(N_c\) values for various \(N_s\) that achieve 95\% of the maximum SE, showing that \(N_c\) does not linearly scale with \(N_s\). This finding is crucial for managing computational complexity at higher \(N_s\) levels, leading to more efficient algorithm designs in systems with increased data streams.

\subsection{Performance Analysis}
\begin{figure}
  \begin{center}
  \includegraphics[width=\columnwidth,height=5.500cm,trim={1.8cm 0cm 0cm 0cm},clip]{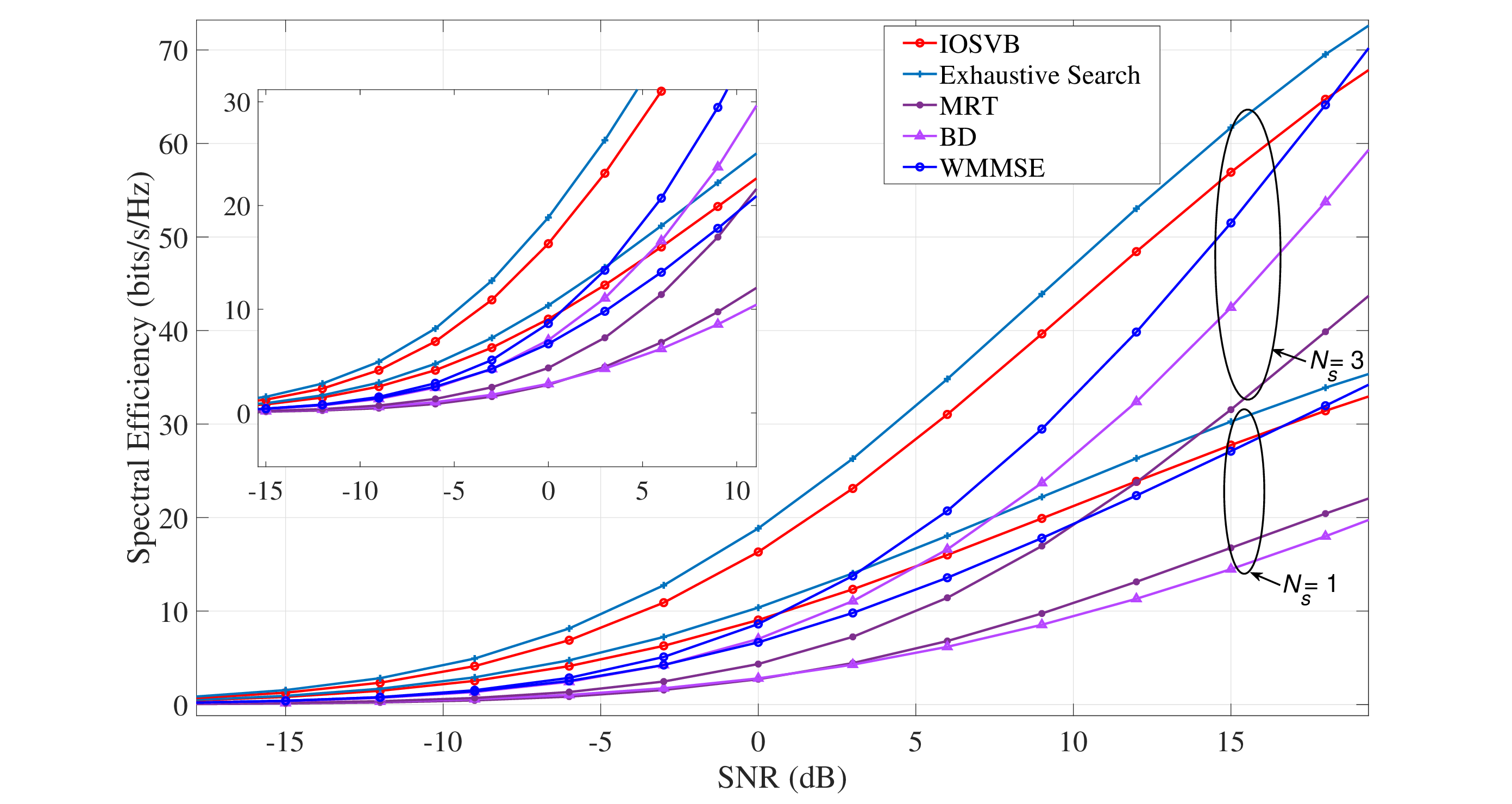}
  \caption{SE comparison of the proposed IOSVB with other existing algorithms for varying SNR levels.}
  \label{SE_comparison}
  \end{center}
\end{figure}

The IOSVB algorithm generates the beamforming matrices by identifying the optimal singular vectors within a reduced search space, effectively reducing multi-user interference and enhancing the SINR. This approach concentrates the computational load on the BS, significantly reducing the processing requirements at the UE side. We conduct an extensive performance evaluation of the IOSVB algorithm against established beamforming techniques: Exhaustive Search, Maximum Ratio Transmission (MRT) \cite{zhang2016mrt}, and Weighted Minimum Mean Square Error (WMMSE) \cite{ren2023mmse}, Block Diagonalization (BD) \cite{spencer2004zero}. The simulations assess SE for a system with \(K = 5\) users, each receiving \(N_s = 1\) and \(N_s = 3\) data streams. Fig. \ref{SE_comparison} shows the SE versus SNR for the four algorithms, averaged over a thousand channel realizations for robustness. In the exhaustive search algorithm, all the beam combinations are explored in a linear search method to determine the beam combination that produces the maximum rate achievable from the system and thus serves as an upper bound in performance comparison.

\begin{table}[ht]
    \caption{$N_c$ value for achieving 95\% of maximum SE}
    \label{selection_Rsel_Ns}
        \centering
    \begin{tabular}{|c|c|}
        \hline
        Data Streams ($N_s$) & Required $N_c$ \\
        \hline
        $2$ & 4 \\
        \hline
        $3$ & 5 \\
        \hline
        $4$ & 5 \\
        \hline
        $5$ & 6 \\
        \hline
    \end{tabular}
\end{table}
The IOSVB algorithm consistently outperforms MRT and WMMSE across typical SNR values (inset in Fig. \ref{SE_comparison}) and closely matches the performance of Exhaustive Search. Notably, IOSVB performs equally well for both \(N_s = 1\) and \(N_s = 3\), demonstrating its versatility and robustness in handling multiple data streams per user. For \(N_s = 3\) and \(N_c = 5\), the times taken by the algorithms to converge are 86.4759 seconds for Exhaustive Search, 0.0059 for IOSVB, 0.0372 seconds for WMMSE, 0.0075 seconds for MRT, and 0.0245 seconds for BD. Note that as the number of users grows, IOSVB’s complexity may increase noticeably. WMMSE, on the other hand, scales more efficiently with user count, providing a potentially lower computational burden in high-user scenarios. MRT prioritizes signal strength over interference mitigation, and its performance degrades significantly when moving from \(N_s = 1\) to \(N_s = 3\), highlighting its limitations in complex MU-MIMO environments. Owing to WMMSE’s theoretical solid foundation in interference management, it performs slightly better under high-SNR conditions.


\section{Conclusion}
While significant research has focused on SU-MIMO beamforming, a notable gap exists in studies dedicated to MU-MIMO beamforming. Current MU-MIMO approaches often fall short in SE due to inadequate interference management. This letter introduced the concept of using the correlation of channel singular vector projections to calculate total interference. Furthermore, we developed IOSVB, a novel beamforming algorithm for MU-MIMO systems designed to identify beams that minimize interference. Through extensive simulations using the NYUSIM channel, we demonstrated that IOSVB outperforms existing algorithms, offering superior performance and reduced computational times. The IOSVB algorithm matches the performance of Exhaustive Search without incurring prohibitive computational costs, making it a promising candidate for next-generation MU-MIMO systems in 6G. Even though the NYUSIM channel is quite realistic for practical settings, imperfect channels obtained from real-world testbeds will be used to evaluate the proposed algorithm's performance in future works. We also intend to perform a thorough performance analysis of the IOSVB under different channel correlation conditions, which will provide more detailed insights.

\appendix

\textit{Corollary 1:} For an orthonormal matrix $\RM{U}_k$, the product $\RM{U}_k^\dag[\RS{\mathcal{I}}_k]\RM{U}_k$ results in a sparse non-square identity matrix, $\RM{I}_k[\RS{\mathcal{I}}_k]$, i.e., $\RM{U}_k^\dag[\RS{\mathcal{I}}_k]\RM{U}_k = \RM{I}_k[\RS{\mathcal{I}}_k]$.

\textit{Proof:} Since $\RM{U}_k$ is orthonormal, $\RM{U}_k \RM{U}_k^\dag=\RM{I}$. The matrix $\RM{U}_k^\dag[\RS{\mathcal{I}}_k]$ is $N_s \times N_r$, with orthonormal rows. Thus, $\RM{U}_k^\dag[\RS{\mathcal{I}}_k]\RM{U}_k$ yields a non-square identity matrix $\RM{I}_k[\RS{\mathcal{I}}_k]$ of size $N_s \times N_r$, containing $N_s$ ones.

 \textit{Lemma 1:} Prove that
 \begin{multline}
||\RM{W}_{k}^\mathrm{\dag} \RM{H}_{k} \left( \sum_{j \neq k} \RM{F}_{j} \RM{F}_{j}^{\dag} \right) \RM{H}_{k}^\mathrm{\dag} \RM{W}_{k}||  
\\
= ||\sum_{j \neq k}
\RM{W}_{k}^\mathrm{\dag} \RM{H}_{k} \RM{F}_{j}
({\RM{W}_{k}^\mathrm{\dag} \RM{H}_{k} \RM{F}_{j}})^\mathrm{\dag}|| .
\label{eqn17}
\ \end{multline}

\textit{Proof:}
By changing the position of the summation, we get
\begin{align}
   & ||\RM{W}_{k}^\mathrm{\dag} \RM{H}_{k} \left( \sum_{j \neq k} \RM{F}_{j} \RM{F}_{j}^{\dag} \right) \RM{H}_{k}^\mathrm{\dag} \RM{W}_{k}||  \nonumber \\
    & =||\sum_{j \neq k}( \RM{W}_{k}^\mathrm{\dag} \RM{H}_{k} \RM{F}_{j}  ) (\RM{F}_{j}^{\dag}  \RM{H}_{k}^\mathrm{\dag} \RM{W}_{k} ) || .
    \label{eq18}
\end{align}
Using the property of a Hermitian matrix in (\ref{eq18}), equation (\ref{eqn17}) can be obtained.

\textit{Lemma 2:} If `$j$' is an interferer, `$k$' is the intended user, 
\begin{equation}
    {|| \RM{W}_k^\dag \RM{H}_k \RM{F}_j||}^2 = {|| \RM{S}_k[\RS{\mathcal{I}}_k,\RS{\mathcal{I}}_k] \RM{V}_k^\dag [\RS{\mathcal{I}}_k]\RM{V}_j[\RS{\mathcal{I}}_j]||}^2
    \label{lema2}
\end{equation}
\textit{Proof:}
Using $\RM{H}_k=\RM{U}_k\RM{S}_k\RM{V}_k^\dag$, we can write
\begin{equation}\label{37}
  {|| \RM{W}_k^\dag \RM{H}_k \RM{F}_j||}^2
 =   {|| \RM{W}_k^\dag \RM{U}_k \RM{S}_k \RM{V}_k^\dag \RM{F}_j||}^2.
\end{equation}
Here, by definition, $\RM{U}_k[\RS{\mathcal{I}}_k]=\RM{W}_k$ and $\RM{V}_j[\RS{\mathcal{I}}_j]=\RM{F}_j.$
Hence,  (\ref{37}) can be rewritten as, 
\begin{equation}
{||\RM{W}_k^\dag \RM{H}_k \RM{F}_j||}^2
 = {|| \RM{U}_{k}^\dag[\RS{\mathcal{I}}_k] \RM{U}_k \RM{S}_k \RM{V}_k^\dag \RM{V}_j[\RS{\mathcal{I}}_j]||}^2.
\end{equation}
Now, using \textit{Corollary 1}, we can write, 
\begin{equation}\label{eq29}
   {||\RM{W}_k^\dag \RM{H}_k \RM{F}_j||}^2 ={|| \RM{I}_k[\RS{\mathcal{I}}_k] \RM{S}_k \RM{V}_k^\dag \RM{V}_j[\RS{\mathcal{I}}_j]||}^2.
\end{equation}
In (\ref{eq29}), $\RM{I}_k[\RS{\mathcal{I}}_k]$ is a matrix of size $N_s \times N_r$ with $N_s$ number of ones that corresponds to indices of $\RS{\mathcal{I}}_k$. Using matrix multiplication, (\ref{eq29}) can be written as  
\begin{equation}
   {||\RM{W}_k^\dag \RM{H}_k \RM{F}_j||}^2 ={||\RM{S}_k[\RS{\mathcal{I}}_k] \RM{V}_k^\dag \RM{V}_j[\RS{\mathcal{I}}_j]||}^2.
\end{equation}
Here, $\RM{S}_k[\RS{\mathcal{I}}_k]$ is a matrix of size $N_s \times N_r$ with $N_s$ non-zero elements, which can also be indicated as $\diag(\sigma_1, \cdots, \sigma_{N_s}).$ Now, we can form a square matrix $\RM{S}_k[\RS{\mathcal{I}}_k,\RS{\mathcal{I}}_k]$ of size $N_s \times N_s$ by extracting the non-zero columns from $\RM{S}_k[\RS{\mathcal{I}}_k].$
Here, the matrix product $\RM{S}_k[\RS{\mathcal{I}}_k] \RM{V}_k^\dag$ is analogous to $\RM{S}_k[\RS{\mathcal{I}}_k,\RS{\mathcal{I}}_k] \RM{V}_k^\dag[\RS{\mathcal{I}}_k]$ as in the latter one we are just disregarding the multiplication with the columns of $\RM{S}_k$ that have only zero values. Thus, using above formulations equation (\ref{lema2}) can be obtained.

\bibliographystyle{IEEEtran}
\bibliography{IEEEabrv,IEEE_Comm_Letters_CL2024_Biblography}
\end{document}